\documentclass[twocolumn,preprintnumbers,notitlepage,superscriptaddress,amsmath,10pt,aps,pra]{revtex4-1}
\usepackage{graphicx}
\usepackage{color}
\usepackage{hyperref}
\usepackage{subfigure}
\usepackage{graphicx}
\usepackage{type1cm}
\usepackage{eso-pic}
\usepackage{color}
\usepackage{amsmath}
\usepackage{amssymb}

\def\blue#1{\textcolor{blue}{#1}}
\def\red#1{\textcolor{black}{#1}}

\begin{document}

\title{Unambiguous quantum state elimination for qubit sequences} %

\preprint{\blue{Draft: \today}} 

\author{Jonathan Crickmore}
\affiliation{SUPA, Institute for Photonics and Quantum Sciences, School of Engineering and Physical Sciences, Heriot-Watt University, Edinburgh EH14 4AS, United Kingdom}
\author{Ittoop V. Puthoor}
\affiliation{SUPA, Institute for Photonics and Quantum Sciences, School of Engineering and Physical Sciences, Heriot-Watt University, Edinburgh EH14 4AS, United Kingdom}
\author{Berke Ricketti}
\affiliation{SUPA, Institute for Photonics and Quantum Sciences, School of Engineering and Physical Sciences, Heriot-Watt University, Edinburgh EH14 4AS, United Kingdom}
\author{Sarah Croke}
\affiliation{SUPA, School of Physics and Astronomy, University of Glasgow, Glasgow G12 8QQ, United Kingdom}
\author{Mark Hillery}
\affiliation{Department of Physics and Astronomy, Hunter College of the City University of New York, 695 Park Avenue, New York, NY 10065 USA, and Graduate Center of the City University of New York, 365 Fifth Avenue, New York, NY 10016}
\author{Erika Andersson}
\affiliation{SUPA, Institute for Photonics and Quantum Sciences, School of Engineering and Physical Sciences, Heriot-Watt University, Edinburgh EH14 4AS, United Kingdom}

\begin{abstract}
Quantum state elimination measurements tell us what states a quantum system does {\em not} have. This is different from state discrimination, where one tries to determine what the state of a quantum system is, rather than what it is not. Apart from being of fundamental interest, quantum state elimination may find uses in quantum communication and quantum cryptography. We consider unambiguous quantum state elimination for two or more qubits, where each qubit can be in one of two possible states. Optimal measurements for eliminating one and two states out of four two-qubit states are given. We also prove that if we want to maximise the average number of eliminated overall $N$-qubit states, then individual measurements on each qubit are optimal. 
\end{abstract}

\maketitle

\section{Introduction}

Quantum measurements more naturally {\em exclude} states rather than indicate precisely which state a quantum system has. For example, if we make a projective measurement in some basis, then we can rule out any state which is orthogonal to the basis state corresponding to the outcome we obtained. 
In spite of this, quantum state elimination~\cite{Stevebook, Caves2002, PBR, SignaturesElim, Bando2014, Wallden2014, Heinosaari1} has not been much explored. More attention has been given to quantum state discrimination~\cite{DiscReview}, where one aims to determine what a quantum state is. 

\red{Quantum state elimination has already generated significant interest, in the sense that the philosophical argument about the reality of the wave function in~\cite{PBR} completely relies on the existence of a particular quantum elimination measurement in an entangled basis.}
Apart from this fundamental interest, quantum state elimination might be useful for applications in quantum information and quantum communication. Examples of this are the \red{communication tasks in~\cite{Perry, Heinosaari2, Havlicek},} and a protocol for quantum oblivious transfer~\cite{RyanOT}. In 1-out-of-2 oblivious transfer, 
a receiver should receive one out of two bits, with no information about the other. The sender should not know which bit was received. 
Among other things, in this paper we investigate unambiguously excluding two out of four possible non-orthogonal two-qubit states. If the four states encode two classical bit values, then such a measurement would tell us either the first bit, the second bit, or the XOR of the two bits. 

\red{In addition to oblivious transfer, one could envisage novel quantum key distribution schemes employing quantum state elimination. Suppose a sender prepares one of four non-orthogonal states, the recipient excludes two of these, and "reconciliation" means that the recipient tells the sender (using an authenticated but not secret classical channel) whether the first, second, or XOR of the bits was received. This bit value then constitutes the secret key, after privacy amplification. As for BB84, eavesdropping can intuitively be detected if the sender and recipient announce some of the bit values (again using an authenticated channel). The novel feature of such a scheme is that the final bit value cannot be thought of as created at the sender and transmitted to the receiver, but is only realised once the receiver's measurement is complete. A similar scheme employing elimination of one among three states has been previously proposed~\cite{Phoenix2000} and optimal eavesdropping attacks shown to have counter-intuitive features. For example the optimal attack to maximise the probability that whenever Alice and Bob agree Eve also knows the shared bit is a unitary~\cite{Flatt2018}. Thus, although quantum key distribution is by now very well-developed, such schemes are conceptually different, and may be of theoretical interest.}

In this paper, we consider unambiguous or ``error-free" quantum state elimination for qubit sequences.
Minimum-error measurements minimize the probability for the result to be wrong, and as a generalisation of this, minimum-cost measurements minimize the average cost associated with the result. Unambiguous measurements, on the other hand, never give a wrong result, which is often possible only if the measurement sometimes fails to give a successful result.
Minimum-cost and minimum-error state elimination are easily seen to be a special case of minimum-cost measurements, with a suitable cost associated with naming a state that was actually prepared.
Moreover, minimum-error state exclusion is equivalent to minimum-error state discrimination among mixtures of the original states~\cite{Bando2014}, and existing results for state discrimination~\cite{DiscReview} therefore apply. 

In contrast, it seems that unambiguous quantum state elimination cannot be usefully re-expressed as unambiguous state discrimination.
This makes it particularly interesting to investigate. 
Here, we consider unambiguous state elimination for sequences of qubits.
Optimal measurements for eliminating one and two states out of four two-qubit states are given. We also prove that if we want to maximise the average number of eliminated overall $N$-qubit states, then individual measurements on each qubit are optimal.

\red{The measurements we will consider will often be not just measurements in entangled basis. They will often be generalised quantum measurements, also called positive operator-valued measurements (POVMs) or probability operator measures (POMs)~\cite{Stevebook, DiscReview}. Similar to projective measurements in some orthonormal basis, generalised measurements are described by a set of Hermitian measurement operators $\Pi_i$, where $i$ labels the outcome. The probability of outcome $i$ is $p_i = {\rm Tr}(\rho \Pi_i)$, if $\rho$ is the state being measured. In contrast to projective measurements, however, the operators $\Pi_i$ need not be projectors. They must nevertheless satisfy
\begin{equation}
\label{eq:genmeasdef}
\Pi_i \ge 0~ \forall~ i, ~~~~\sum_i \Pi_i = 1.
\end{equation}
The first of these conditions means that all eigenvalues of $\Pi_i$ should be positive. This is necessary because probabilities should be nonnegative numbers. The second condition means that the sum of the probabilities for all outcomes, including the measurement failing to give the desired information, should be equal to one. That is, although the measurement is not a projection in an orthonormal basis, the outcomes are nevertheless mutually exclusive in the sense that if one outcome occurs, another one does not occur.}

Generalizing the situation in \cite{PBR}, we will consider states of two or more qubits. Each qubit has one of the states
\begin{equation}
|\pm\theta\rangle = \cos\theta|0\rangle \pm \sin\theta |1\rangle
\end{equation} 
where $0\le\theta\le 45^\circ$. 
The single-qubit states orthogonal to $|\pm\theta\rangle$ are denoted by 
\begin{equation}
|\overline{\pm\theta}\rangle = \sin\theta|0\rangle \mp \cos\theta |1\rangle.
\end{equation} 
For two qubits, the four possible states are
\begin{eqnarray}
\label{eq:fourstates}
|+\theta, +\theta\rangle &=& \cos^2\theta |00\rangle + \sin^2\theta|11\rangle + \cos\theta\sin\theta ( |01\rangle + |10\rangle ) \nonumber\\
|+\theta, -\theta\rangle &=& \cos^2\theta |00\rangle - \sin^2\theta|11\rangle - \cos\theta\sin\theta (|01\rangle - |10\rangle )\nonumber\\
|-\theta, +\theta\rangle &=& \cos^2\theta |00\rangle - \sin^2\theta|11\rangle + \cos\theta\sin\theta (|01\rangle - |10\rangle )\nonumber\\
|-\theta, -\theta\rangle &=& \cos^2\theta |00\rangle + \sin^2\theta|11\rangle - \cos\theta\sin\theta (|01\rangle + |10\rangle ).\nonumber\\
\end{eqnarray}
Pusey, Barrett and Rudolph~\cite{PBR} mainly considered the scenario with two qubits and $2\theta = 45^\circ$, when a measurement in an entangled basis can unambiguously eliminate one of the four possible states with success probability equal to 1. We will call this the ``PBR measurement".

If we for clarity take the two states for each qubit to be $|0\rangle$ and $|+\rangle=(|0\rangle+|1\rangle)/\sqrt 2$ instead of $|\pm\theta\rangle$, then the PBR measurement basis is given by
\begin{eqnarray}
\label{eq:PBR}
|\neg(0,0)\rangle&=&\frac{1}{\sqrt 2}\left(|0,1\rangle+ |1,0\rangle\right)\nonumber\\
|\neg(0,+)\rangle&=&\frac{1}{\sqrt 2}\left(|0,-\rangle+ |1,+\rangle\right)\nonumber\\
|\neg(+,0)\rangle&=&\frac{1}{\sqrt 2}\left(|+,1\rangle+ |-,0\rangle\right)\nonumber\\
|\neg(+,+)\rangle&=&\frac{1}{\sqrt 2}\left(|+,-\rangle +\red{|-,+\rangle}\right).
\end{eqnarray}
 If the result $\neg(00)$ is obtained, the state cannot have been $|00\rangle$, since $|\neg(00)\rangle \perp |00\rangle$, and similarly for the other results.

If one wants to unambiguously eliminate states, but only local measurements on each qubit are allowed, then it is best to perform optimal unambiguous state discrimination on each qubit. Two equiprobable states $|\phi\rangle$ and $|\theta\rangle$ can be unambiguousuly distinguished with the probability $1-|\langle\phi|\theta\rangle|$~\cite{Ivan, Dieks, Peres}, meaning that $|\pm \theta\rangle$ can be distinguished from each other with probability $1-p_f=1-\cos(2\theta)$. One fails to eliminate any state at all iff both measurements fail. This happens with probability $p_f^2$, which is equal to 1/2 for $2\theta=45^\circ$. That is, even if there are no correlations between the two qubits, a measurement in an entangled basis can exclude a single two-qubit state more often than local measurements.

\section{Eliminating one out of four states}

If $45^\circ<2\theta\le 90^\circ$, one can construct a measurement that always eliminates one state for example as follows. First, couple each ``system" qubit to an ancilla $|\phi\rangle_a$, using a suitable unitary transform $V$ so that $V|\pm\theta\rangle\otimes|\phi\rangle_a = |\pm 22.5^\circ\rangle\otimes |\phi_\pm\rangle_a$. This is possible if we choose $\langle\phi_-|\phi_+\rangle = \sqrt 2 \langle -\theta|\theta\rangle$, which is always possible if $45^\circ \le 2\theta\le 90^\circ$. Then, proceed to make the PBR measurement on the ``system" qubits. (The measurement basis for $|\pm 22.5^\circ\rangle$ is constructed analogously to that for $|0\rangle, |+\rangle$ in \eqref{eq:PBR}.) The ancilla can be discarded, or can sometimes be used to unambiguously eliminate more states~\cite{TBS}. %

When $\red{0^\circ}\le 2\theta<45^\circ$, it is impossible to always eliminate one state. To start with, it can be shown that when ruling out one of four \red{equiprobable} two-qubit states, the \red{optimal} measurement operators corresponding to successful outcomes obey the same symmetry as the states. 
The four two-qubit states are transformed into each other when $U$ is applied to the first, second, or both qubits, where $U=|0\rangle\langle 0| - |1\rangle\langle 1|$, with $U^2=1$. 
Then there exists an optimal measurement with the same symmetry,
\begin{eqnarray}
\label{eq:symmetry}
\Pi_{\neg(-+)} &=& (U_1 \otimes I_2)\Pi_{\neg(++)}(U_1^\dagger\otimes I_2),\nonumber\\
\Pi_{\neg(+-)} &=& (I_1 \otimes U_2)\Pi_{\neg(++)}(I_1\otimes U_2^\dagger),\nonumber\\
\Pi_{\neg(--)} &=& (U_1 \otimes U_2)\Pi_{\neg(++)}(U_1^\dagger\otimes U_2^\dagger),
\end{eqnarray}
where $U_i=|0\rangle\langle 0|-|1\rangle\langle 1|$ applied to the $i^{th}$ qubit, and $I_i$ is the identity operation for the $i^{th}$ qubit, \red{and we have used the shorthand notation``$++$" for the state $|\theta,\theta\rangle$, ``$+-$" for $|\theta, -\theta\rangle$ and so on.}  To see this, suppose that an optimal measurement strategy has the measurement operators $\widetilde\Pi_{\neg(++)}, \widetilde\Pi_{\neg(+-)}, \widetilde\Pi_{\neg(-+)}, \widetilde\Pi_{\neg(- -)}$ (in addition to a possible failure measurement operator $\widetilde\Pi_{f,1}$), but that it does not have the symmetry in \eqref{eq:symmetry}. The measurement with operators
\begin{eqnarray}
\Pi_{\neg(++)} &=& \frac{1}{4}\left(\widetilde\Pi_{\neg(++)} +U_1 \widetilde\Pi_{\neg(-+)}U_1^\dagger+ U_2 \widetilde\Pi_{\neg(\red{+-})}U_2^\dagger \right. \nonumber\\
&&\left.+U_1\otimes U_2 \widetilde\Pi_{\neg(- -)}U_1^\dagger\otimes U_2^\dagger\right)
\end{eqnarray}
and analogously for the other measurement operators, is easily seen to obey the symmetry. \red{If the four states are equiprobable, it will  have the same success probability as the optimal measurement we started from. To show this, let us examine the failure operator for the symmetrized measurement, 
\begin{eqnarray}
\Pi_{f,1} &=& \frac{1}{4}\left(\widetilde\Pi_{f,1}+U_1\widetilde\Pi_{f,1}U_1^\dagger + U_2 \widetilde\Pi_{f,1} U_2^\dagger\right.\nonumber\\
&& + \left.U_1 \otimes U_2 \widetilde\Pi_{f,1}f U_1^\dagger\otimes U_2^\dagger
\right).
\end{eqnarray}}
Let us also note that the operation $S(\Pi ) = \frac{1}{2}\red{(\Pi + U\Pi U^\dagger)}$, where $U=|0\rangle\langle 0|-|1\rangle\langle 1|$, deletes the off-diagonal elements of a $2\times 2$ operator (or density matrix) in the $\{|0\rangle, |1\rangle\}$ basis. This also means that the ``symmetrized" $\Pi_{f,1}$
is diagonal in the $\{|00\rangle, |01\rangle, |10 \rangle, |11\rangle\}$ basis.

\red{The failure probability for the measurement we started with is given by
\begin{eqnarray}
\widetilde p_{f,1}&=&p_{++}p(\widetilde f|++) + p_{+-}p(\widetilde f|+-)+ p_{-+}p(\widetilde f|-+)\nonumber\\
&&+p_{--}p(\widetilde f|- -),
\end{eqnarray}
where $p_{++}$ is the prior probability for state $|\theta,\theta\rangle$ to occur, and $p(\widetilde f|++)$ is the conditional probability for failing to exclude a state, given that the state was $|\theta,\theta\rangle$, and analogously for the other three states. The failure probability for the symmetrised measurement in \eqref{eq:symmetry} is then
\begin{eqnarray}
p_{f,1}&=&\frac{1}{4}(p_{++}+p_{+-}+p_{-+}+p_{- -}) [p(\widetilde f|++)\nonumber\\
&&+p(\widetilde f|+-)+p(\widetilde f|-+)+p(\widetilde f|- -)]\\
&=&p(\widetilde f|++)+p(\widetilde f|+-)+p(\widetilde f|-+)+p(\widetilde f|- -).\nonumber
\end{eqnarray}
If all four states are equiprobable (or if all four conditional failure probabilities are equal), then this is equal to the failure probability $\widetilde p_{f,1}$ of the measurement we started with.}

Thus we can without loss of generality assume that an optimal measurement obeys the symmetry, \red{if the states are equiprobable and} if we allow mixed measurement operators. (A similar construction is possible also in other  cases where there is some sort of symmetry among the possible states.)

It can further be shown that in the range $0\le2\theta <45^\circ$, the optimal measurement obeying the symmetry has $\Pi_{\neg(++)}=|X\rangle\langle X|$ for a particular $|X\rangle$, meaning that the other measurement operators also are proportional to pure-state projectors.
Consider a strategy which obeys the symmetry but might have mixed measurement operators, with
\begin{equation}
\label{eq:mixedPi}
\Pi_{\neg(++)} = \sum_{i=1}^3 |X_i\rangle\langle X_i|,
\end{equation}
where
\begin{equation}
|X_i\rangle = \sum_{j,k=0,1} c^i_{jk} |jk\rangle
\end{equation} 
are proportional to the eigenstates of $\Pi_{\neg(++)}$ (we have absorbed the eigenvalues of $\Pi_{\neg(++)}$ into the states in \eqref{eq:mixedPi}). 
The condition $\langle \theta,\theta|\Pi_{\neg(++)}|\theta,\theta\rangle=0$, that is, we never incorrectly eliminate a state, reads
\begin{equation}
\label{eq:overlapmixed}
c^i_{00}\cos^2\theta+(c^i_{01}+c^i_{10})\sin\theta\cos\theta + c^i_{11}\sin^2\theta=0
\end{equation}
for $i=1,2,3$.
The failure operator is 
\begin{equation}
\Pi_{f,1} = I-\sum_{jk=+,-}\Pi_{\neg(jk)} = I-4\sum_{j,k=0,1}\sum_{i=1}^3 |c^i_{jk}|^2|jk\rangle\langle jk|.
\end{equation}
Since $\Pi_{f,1}\ge 0$,
it must hold that $\sum_{i=1}^3|c^i_{jk}|^2\le 1/4$.
If the states are equally likely, then the failure probability is
\begin{eqnarray}
p_{f,1} &=& \frac{1}{4}\sum_{j,k=\pm\theta}\langle j,k|\Pi_{\red {f, 1}}|j,k\rangle \nonumber \\
&=& 1-4\sum_{i=1}^3\left[|c^i_{00}|^2\cos^4\theta + \right. \\ 
&& (|c^i_{01}|^2+|c^i_{10}|^2) \sin^2\theta\cos^2\theta 
\left.+|c^i_{11}|^2\sin^4\theta\right]. \nonumber
\end{eqnarray}
We want to minimize $p_{f,1}$, which means that we want to maximise the sum in the last two lines in the above equation, subject to \eqref{eq:overlapmixed}, $\sum_{i=1}^3|c^i_{jk}|^2\le 1/4$, and with $|X_i\rangle$ orthogonal to each other. The coefficient multiplying $\sum_{i=1}^3|c_{00}^i|^2$  is the largest. Writing \eqref{eq:overlapmixed} as
\begin{equation}
\label{eq:overlapc00}
c^i_{00}=-(c^i_{01}+c^i_{10})\tan\theta - c^i_{11}\tan^2\theta,
\end{equation}
and taking into account that for $0\le 2\theta < 45^\circ$, it holds that $0\le 2 \tan\theta + \tan^2\theta \le 1$, one realizes that in order to maximise the success probability, it is optimal to choose $\Pi_{\neg(++)} = |X_1\rangle\langle X_1|$, with  $c^1_{01}=c^1_{10}=c^1_{11} = -1/2$ and $c^1_{00} = \tan\theta(1+\frac{1}{2}\tan\theta)$, and all other $c^i_{jk}=0$. The failure operator is then
\begin{equation}
\Pi_{f,1}=[1-\tan^2\theta(2+\tan\theta)^2]|0\rangle\langle 0|,
\end{equation}
and the failure probability is
\begin{eqnarray}
p_{f,1}&=&[\cos(2\theta) - \sin(2\theta)][1+\sin(2\theta)].
\end{eqnarray}

\section{Eliminating two out of four states}

It turns out that if $\theta$ is large enough, then it becomes possible to eliminate not just one, but two out of the four two-qubit states. As we will show, this occurs if $\cos (2\theta) \le \sqrt{2} -1$, meaning that $2\theta \gtrsim 65.5^\circ$.
There are six ways to choose two states out of four. 
Using the previous shorthand notation``$++$" for the state $|\theta,\theta\rangle$, ``$+-$" for $|\theta, -\theta\rangle$ and so on, the pairs we can eliminate are 
\begin{eqnarray}
\label{eq:pairs}
&\{++,+-\}, ~\{++, -+\}, ~\{+-, --\}, ~\{-+,--\}, & \nonumber \\
&\{+-, -+\}, ~\{++, --\}.&
\end{eqnarray}
We will label these pairs with the letters $A$ to $F$, in the order they appear above. In the four first pairs, $A$ -- $D$, the state of either the first or the second qubit is the same in both excluded states. That is, eliminating one of the first pairs means determining the state of one of the qubits.
For the two last pairs, $E$ and $F$, we have eliminated the possibility that the two qubits have the same state, or that they have different states.

First, let's bound for what range of $\theta$ it is possible to unambiguously eliminate two states with success probability 1. 
Suppose that the prepared state really is $|\theta, \theta\rangle$, but that we only know that it is either $|\theta, \theta\rangle$ or $|-\theta, -\theta\rangle$, \red{with equal prior probabilities}. The possible outcomes are now $\{+-,--\}, \{-+, --\}$ and $\{+-, -+\}$, with labels $C,D,E$, plus the ``failure" outcome, if it occurs. If the outcome is $C$ or $D$, that is, $\{+-, --\}$ or $\{-+,--\}$, then we have succeeded in unambiguously telling $|\theta, \theta\rangle$ from $|-\theta, -\theta\rangle$. This must occur with a probability at most $1-\langle -\theta,-\theta|\theta,\theta\rangle=1-\cos^2(2 \theta)$, meaning that $p(C|++)+p(D|++)\le 1-\cos^2(2 \theta)$ \red{(the conditional success probabilities for optimally distinguishing between two equiprobable states are also equal for both states)}. Now, suppose that the prepared state still is $|\theta, \theta\rangle$, but that we only know that it is either $|\theta, \theta\rangle$ or $|\theta, -\theta\rangle$, \red{again with equal prior probabilities.}. If the outcome is  $C$ or $E$, $\{+-,--\}$ or $\{+-,-+\}$, then we have succeeded in distinguishing $|\theta,\theta\rangle$ and $|\theta,-\theta\rangle$, which can occur with probability at most $1-\langle\red{\theta,-\theta}|\theta,\theta\rangle=1-\cos(2\theta)$, giving $p(C|++)+p(E|++)\le 1-\cos(2\theta)$. Similarly, if \red{we know that} the state is either $|\theta,\theta\rangle$ or $|-\theta,\theta\rangle$, \red{ with equal prior probabilities}, the outcomes $D$ and $E$, $\{-+,--\}$ and $\{+-,-+\}$, can occur with probability at most probability $1-\cos(2\theta)$, giving $p(D|++)+p(E|++)\le 1-\cos(2\theta)$. Adding these three inequalities, and assuming that the failure probability
is zero, so that $p(C|++)+p(D|++)+p(E|++)=1$,
we obtain
\begin{equation}
2\le 3-2\cos(2\theta)-\cos^2(2\theta)=
4-[1+\cos(2\theta)]^2,
\end{equation}
meaning that $\cos(2\theta)\le\sqrt 2-1$ must hold if the measurement never fails.

This bound turns out to be tight. 
We will now construct a measurement that unambiguously eliminates pairs of states. If  $\cos(2\theta)\le\sqrt 2-1$, the success probability is equal to one, otherwise it is less than one. Some intuitive assumptions will be made, but 
\red{by comparison with results in Sec. \ref{sec:ave}, in particular Eq. \eqref{eq:twoave}}, it can be seen that the constructed measurement is optimal also when $\cos(2\theta) > \sqrt 2-1$. 

First, look at $E$ and $F$, the two last cases in \eqref{eq:pairs}. An orthogonal (but unnormalised) basis for the space spanned by the states $|\theta, \theta\rangle$ and $|-\theta, -\theta\rangle$ is given by 
\begin{equation}
\label{eq:samestates}
\{\cos^2\theta|00\rangle+\sin^2\theta|11\rangle, ~~|01\rangle + |10\rangle\}.
\end{equation} 
Similarly, an orthogonal (but unnormalised) basis for the space spanned by the states $|\theta, -\theta\rangle$ and $|-\theta, \theta\rangle$ is
\begin{equation}
\label{eq:diffstates}
\{\cos^2\theta|00\rangle-\sin^2\theta|11\rangle, ~~|01\rangle - |10\rangle\}.
\end{equation}
Therefore, if some measurement operators are proportional to projectors onto $|\psi^\pm_{01}\rangle=|01\rangle \pm |10\rangle$ and $|\psi^\pm_{cs}\rangle=\sin^2\theta|00\rangle\pm\cos^2\theta|11\rangle$, then these outcomes would allow us to unambiguously rule out either $\{+-, -+\}$ or $\{++, --\}$. (More generally, such measurement operators could also be proportional to projectors onto superpositions of $|\psi^\pm_{01}\rangle$ and $|\psi^\pm_{cs}\rangle$, and mixtures of such projectors.)

The sum of all measurement operators, including a possible failure operator, must be the identity operator. If projectors onto the unnormalized states $|01\rangle \pm |10\rangle$ are included among the measurement operators, with the same weight $\alpha$, then the resulting contribution to the sum of all measurement operators is
$2\alpha(|01\rangle\langle 01|+ |10\rangle\langle 10|)$. 
Similarly, if projectors onto the states $\sin^2\theta|00\rangle\pm\cos^2\theta|11\rangle$ are included among the measurement operators, with the same weight $\beta$, then the contribution to the sum of all measurement operators is 
\begin{equation}
\label{eq:twoproj}
2\beta(\sin^4\theta |00\rangle\langle 00|+ \cos^4\theta|11\rangle\langle 11|).
\end{equation} 
It holds that $\sin^4\theta \le \cos^4\theta$ if $0 \le\theta \le 45^\circ$. Therefore, to obtain a measurement that {\em always} eliminates one pair of states, other measurement operators would need to make a greater contribution to $|00\rangle\langle 00|$ than to $|11\rangle\langle11|$. 

Let us then consider the first four pairs, $A$ -- $D$, in \eqref{eq:pairs}. The states in the pair $\{++,+-\}$ span the space $|\theta\rangle\langle\theta|\otimes I_2$, where $I_2$ is the identity operator on the second qubit. 
A measurement operator that corresponds to ruling out $\{++,+-\}$ must therefore have the form $|\overline{\theta}\rangle_{11}\langle \overline{\theta}|\otimes \pi_2$, where $\pi_2$ is some operator acting on the second qubit. We will choose $\pi_2\propto |0\rangle_{22}\langle 0|$, since we need to make up a ``shortfall" in $|00\rangle\langle 00|$.
Similarly, ruling out the pair $\{++, -+\}$ must correspond to a measurement operator $\pi_1\otimes|\overline{\theta}\rangle_{22}\langle \overline{\theta}|$, and we will again choose $\pi_1\propto |0\rangle_{11}\langle 0|$. Finally, ruling out the remaining two pairs will correspond to measurement operators proportional to $|\overline{-\theta},0\rangle\langle \overline{-\theta}, 0|$ and $|0,\overline{-\theta}\rangle\langle 0,\overline{-\theta}|$.
It holds that
\begin{equation}
|\overline{\theta}\rangle\langle \overline{\theta}| + |\overline{-\theta}\rangle\langle \overline{-\theta}|=2\sin^2\theta|0\rangle\langle 0| + 2\cos^2\theta |1\rangle\langle 1|.
\end{equation}
Hence, if projectors onto the four states $|\overline{\theta}, 0\rangle$, $|\overline{-\theta}, 0\rangle$, $|0,\overline{\theta}\rangle$, $|0, \overline{-\theta}\rangle$ are all included among the measurement operators with the same weight $\gamma$, then the resulting contribution to the sum of all measurement operators is
\begin{equation}
\label{eq:fourproj}
\gamma[4\sin^2\theta |00\rangle\langle 00| + 2 \cos^2\theta(|01\rangle\langle 01| + |10\rangle\langle 10 |)].
\end{equation}

Collecting all contributions to the sum of all measurement operators, the sum is less than or equal to the identity operator if the inequalities
\begin{eqnarray}
\label{eq:sumineq}
2\beta\sin^4\theta+4\gamma\sin^2\theta\le 1\nonumber\\
2 \beta\cos^4\theta\le 1\nonumber\\
2\alpha+2\gamma\cos^2\theta\le 1
\end{eqnarray}
are all satisfied. If equality can be reached in all three inequalities, then the measurement always succeeds in eliminating two states.
Evidently, we should choose $\beta = 1/(2 \cos^4\theta)$. If the angle $\theta$ is large enough, so that $\cos 2\theta \le \sqrt{2}-1$, then we can satisfy all three inequalities by choosing
\begin{equation}
\gamma = \frac{1-\tan^4\theta}{4\sin^2\theta}, ~~\alpha = \frac{1}{2}-\gamma\cos^2\theta.
\end{equation}
If $\cos 2\theta > \sqrt{2}-1$, then we must instead choose $\gamma=1/(2\cos^2\theta)$ and $\alpha=0$ so as not to violate the last inequality in \eqref{eq:sumineq}. In this case, there will be a failure operator proportional to $|00\rangle\langle 00|$,
\begin{eqnarray}
\Pi_{f,2}
&=&[2-(1+\tan^2\theta)^2]|00\rangle\langle00|.
\end{eqnarray}
The failure probability is 
\begin{equation}
\label{eq:failtwo}
p_{f, 2}=\cos^4\theta[2-(1+\tan^2\theta)^2] = 2\cos^4\theta-1.
\end{equation}
If $\cos^2\theta = 1/\sqrt{2}$, then $p_{f, 2}=0$ as it should. 

The measurement operators for outcomes $A$ -- $F$ are given by
\begin{eqnarray}
\Pi_A &=& \gamma |\overline{\theta},0\rangle\langle \overline{\theta},0|, 
~~~~~~\Pi_B = \gamma |0,\overline{\theta}\rangle\langle 0, \overline{\theta}|,\nonumber\\
\red{ \Pi_C} &=& \gamma |0,\overline{-\theta}\rangle\langle 0, \overline{-\theta}| ,
~~\red{ \Pi_D }= \gamma |\overline{-\theta},0\rangle\langle \overline{-\theta},0|,\nonumber\\
\Pi_E &=& \alpha|\psi^+_{01}\rangle\langle\psi^+_{01}|+\beta |\psi^+_{cs}\rangle\langle\psi^+_{cs}|,\nonumber\\
\Pi_F &=& \alpha|\psi^-_{01}\rangle\langle\psi^-_{01}|+\beta |\psi^-_{cs}\rangle\langle\psi^-_{cs}|,
\end{eqnarray}
with $\alpha, \beta,\gamma$ given above. In the range $\cos (2\theta)\le \sqrt{2}-1$, where the success probablility is equal to 1, it holds that
\begin{equation}
p_A=p_B=p_C=p_D=\frac{1}{2}\cos(2\theta), ~~p_E=p_F=\frac{1}{2}-\cos(2\theta),
\end{equation}
and in the range $\cos (2\theta) > \sqrt{2}-1$, when the success probability is less than 1, it holds that
\begin{equation}
p_A=p_B=p_C=p_D=\sin^2\theta\cos^2\theta, ~~p_E=p_F=\sin^4\theta.
\end{equation}
We conjecture that this measurement is optimal for unambiguously eliminating two out of four states also in the range when we cannot do so 100\% of the time. 

If we rule out that the states of a number of quantum systems are the same, then they must be different, and vice versa. Quantum state elimination is therefore related to quantum state comparison~\cite{Stevecomp, Comp1, Comp2}. As shown in \cite{Stevecomp}, the optimal success probability for unambiguously determining if two states, each drawn from the set $\{|\theta\rangle, |-\theta\rangle\}$, are the same or different, is $1-\cos (2\theta)=2\sin^2\theta$. (The probability for the outcomes ``same" and ``different" are each half of this.) This is consistent with what we have found here. For the range $\cos(2\theta) \le \sqrt{2}-1$ where two states are always ruled out, it holds that $p_E+p_F=1-2\cos(2\theta )$. For the range $\cos(2\theta) > \sqrt{2}-1$, it holds that $p_E+p_F=2\sin^4\theta$. That is, if we are only interested in whether the states are the same or different, outcomes $E$ and $F$, then it is possible to obtain these outcomes somewhat more often than when maximising $p_A+p_B+p_C+p_D+p_E+p_F$.

\section{Maximising the average number of eliminated states}
\label{sec:ave}

Let us now consider $N$ qubits, each in the state $|\theta\rangle$ or $|-\theta\rangle$. We will consider a measurement that sometimes eliminates 0, 1, 2 states, and so on, all the way to eliminating all but one of the $2^N$ states.
We will show that if one wants to maximise the average number of eliminated states, then unambiguous measurements on each qubit individually are optimal. First, let's see how many states are excluded in this case. The failure probability for each single-qubit measurement  is $p_f=|\langle-\theta|\theta\rangle| = \cos(2\theta)$.
The probability that $M$ measurements fail and $N-M$ measurements succeed is then ${N\choose M}p_f^{M}(1-p_f)^{N-M}$. If $M$ measurements fail, then there are $2^M$ possible states left, and the average number of states eliminated is
\begin{eqnarray}
\label{eq:unambave}
\langle S \rangle &=&2^N-\sum_{M=0}^N{N\choose M}p_f^{M}(1-p_f)^{N-M}2^M\\
 &=& 2^N -[2p_f+(1-p_f)]^N=2^N-(1+p_f)^N.\nonumber
\end{eqnarray}
In other words, each single-qubit measurement leaves on average $1+p_f$ possible states for that qubit, so that the average number of possible states after making all single-qubit measurements is $(1+p_f)^N$. If $N=2$, individual measurements eliminate on average 1.086 states, which is slightly more than the PBR measurement which always eliminates 1 state. We can also easily calculate e.g. for what combinations of $N$ and $\theta$ half of all states will be eliminated on average.

To bound the average number of states that any measurement could unambiguously eliminate, consider using such a measurement to distinguish between the states $|x\rangle$ and $|y\rangle$, where $x$ and $y\in\{0, 2^N-1\}$ will label two $N$-qubit states which now occur with probability 1/2 each. That is, we will use the elimination measurement for another purpose than what it might have been optimised for. 
If we exclude $|x\rangle$, then the state must have been $|y\rangle$, and vice versa. No measurement can distinguish between 
the equiprobable states $|x\rangle$ and $|y\rangle$ more often than with probability $1-|\langle x|y\rangle|=1-p_f^M$, if $x$ and $y$ differ from each other in $M$ positions, where $p_f=\cos(2\theta)$ as before. For any unambiguous elimination measurement it must therefore hold that
\begin{eqnarray}
\label{eq:xyineq}
p(\neg x|y)p(y) + p(\neg y|x)p(x) &=& \frac{1}{2}[p(\neg x|y) + p(\neg y|x)] \nonumber\\ &\le& 1-p_f^M,
\end{eqnarray}
where the LHS in the equality gives the probability for the elimination measurement to succeed in distinguishing between $|x\rangle$ and $|y\rangle$, the RHS is the optimal success probability, and the prior probabilities for the two states to occur are $p(x)=p(y)=1/2$. 

We will now sum the inequality \eqref{eq:xyineq} over all possible $x$ and $y$ (and multiply the sum with $1/2^N$). To start with, for the LHS in this sum of inequalities it holds that
\begin{equation}
\sum_{x, y=0}^{2^N-1} \frac{1}{2}[p(\neg x|y) + p(\neg y|x)] = \sum_{x, y=0}^{2^N-1} p(\neg x|y). 
\end{equation}
(Excluding $x=y$ from the sum would make no difference, since we are only considering unambiguous elimination measurements for which $p(\neg x|x)=0$, and if $x=y$, then $p_f^M=1$ since $x$ and $y$ differ in no positions.)
Moreover, we can write the conditional probability $p(\neg x|y)$ as a sum over probabilities to exclude different sets of states, where each set in this sum contains the state $|x\rangle$. We will denote the possible sets of states by $S_{K,j}$, where $K$ is the number of $N$-qubit states contained in the set, and $j$ labels the different sets of the same size. \red{Excluding the states in a set $S_{K,j}$ means excluding {\em only} these states, no more and no less. The outcomes corresponding to excluding different sets $S_{K,j}$ are in this sense mutually exclusive. Also, the sum of all the probabilities for excluding different sets -- including the empty one -- will be equal to one, as stated in \eqref{eq:genmeasdef}.}
It then holds that
\begin{equation}
\label{eq:Kjsum}
p(\neg x|y) = \sum_{K=1}^{2^N-1} \sum_{j:|x\rangle\in S_{K,j}} p(\neg S_{K,j}|y),
\end{equation}
where $p(\neg S_{K,j}|y)$ is the probability to exclude the set $S_{K,j}$, given that the state was $|y\rangle$, and we are summing over $j$ such that $S_{K,j}$ contains $|x\rangle$.
If we now sum \eqref{eq:Kjsum} over $x$, then each set $S_{K,j}$ will occur in this sum $K$ times, once for each state it contains. It therefore holds that
\begin{equation}
\label{eq:Ksum}
\sum_{x=\red{0}}^{2^N-1}p(\neg x|y) = \sum_{K=1}^{2^N-1} K p(\neg K|y),
\end{equation}
where 
\begin{equation}
p(\neg K|y)=\sum_j p(\neg S_{K,j}|y)
\end{equation}
 is the probability to exclude $K$ states, given that the state was $y$. Now, if we use the same state elimination measurement in a different scenario, where each state $|y\rangle$ occurs with the same probability, $p(y)=1/2^N$, then it holds that
 \begin{eqnarray}
 \frac{1}{2^N} \sum_{x,y=0}^{2^N-1}p(\neg x|y) &=& \sum_{K=1}^{2^N-1} \sum_{y=0}^{2^N-1}K p(\neg K|y)p(y)\nonumber\\
&=&\sum_{K=1}^{2^N-1} K p(\neg K). 
 \end{eqnarray}
This is equal to the average number of states eliminated when each $N$-qubit state $|y\rangle$ occurs with the same probability.

On the RHS of the sum of inequality \eqref{eq:xyineq} over $x$ and $y$,
$p_f$ appears with different powers. For each $|y\rangle$, there are ${N\choose M}$ different  states $|x\rangle$ that differ from $|y\rangle$ in $M$ positions, and this will determine the coefficient of each $p_f^M$. For a given $y$ it therefore holds that 
\begin{equation}
\sum_{x=0}^{2^N-1} 1-p_f^M = 2^N-\sum_{M=0}^N{N \choose M} p_f^M = 2^N-(1+p_f)^N.
\end{equation}
Combining our results for the LHS and RHS for the sum over $x$ and $y$ of \eqref{eq:xyineq}, multiplied by $1/2^N$, we obtain
\begin{eqnarray}
\label{eq:avebound}
\sum_{K=1}^{2^N-1} K p(\neg K) &\le&
2^N-(1+p_f)^N. 
\end{eqnarray}
That is, we have obtained a bound on how many states can be eliminated on average, by any measurement, when each $N$-qubit state appears equally often. As stated above, the LHS equals the average number of states which are eliminated. The RHS, by comparison with \eqref{eq:unambave},  is precisely the number of states that will be eliminated by individual unambiguous measurements. It follows that  separate unambiguous measurements on each qubit are optimal; this eliminates the highest possible number of states on average.

Equation \eqref{eq:avebound} can also be used to bound how often measurements that eliminate certain numbers of states can succeed. It is possible that ``entangled" measurements eliminate as many states on average as separate unambiguous measurements do, but they cannot do better. For example, consider two qubits with four states in total. It is possible to increase the probability to eliminate two of these states, at the expense of lowering the probabilities to eliminate one or three states. But it still has to hold that
\begin{equation}
\label{eq:twoave}
2p(\neg 2)\le 4-(1+p_f)^2 \Leftrightarrow p_f \le \sqrt{4-2p(\neg 2)}-1.
\end{equation}
Comparing with \eqref{eq:failtwo}, with $p_f=\cos (2\theta)$, we see that this bound is tight. That is, the optimal entangled six-outcome measurement that eliminates a pair of states as often as possible also eliminates as many states as possible on average.

In general, however, similar bounds do not have to be tight. If we for example consider elimination of one state, again for two qubits, then we obtain
\begin{equation}
\red{p(\neg 1) \le 4-(1+p_f)^2\Leftrightarrow p_f\le \sqrt{4-p(\neg 1)}-1.}
\end{equation}
If $p(\neg 1)=1$, we obtain $p_f\le \sqrt 3 -1\approx 0.732$. However, we know that it is possible to deterministically eliminate one state for $45^\circ\le\red{ 2}\theta\le 90^\circ$, and $\cos(45^\circ)=1/\sqrt 2 \approx 0.707$, so that the bound from the average number of eliminated states  is not tight in this case. Nevertheless, we can generally say that for the probability for eliminating $K$ states, among $N$ qubits, it must hold that
\begin{equation}
Kp(\neg K)\le 2^N-(1+p_f)^N.
\end{equation}
If we take $K=2^N-1$, which corresponds to unambiguous discrimination among all $2^N$ states, where the optimal success probability is known to be $p(\neg (2^N-1))=(1-p_f)^N$ (the optimal measurement is to perform unambiguous discrimination individually on each qubit, and the overall measurement succeeds if all measurement succeed), we obtain
\begin{equation}
\label{eq:discineq}
(2^N-1)(1-p_f)^N\le 2^N-(1+p_f)^N.
\end{equation}
This inequality is not tight \red{for $N\ge 2$} unless $p_f=0$ \red{or $p_f=1$}, that is, the single-qubit states are orthogonal \red{or identical}. \red{The inequality is also tight for $N=1$, for any $p_f$, as should be expected; if we consider only one qubit, then the optimal measurement that excludes all but one state of ``all qubits" (a single one) is by definition the optimal unambiguous measurement for a single qubit.}

\red{To show that the relation otherwise is not tight, consider the function
\begin{equation}
g(p_f) = 2^N-(1+p_f)^N-(2^N-1)(1-p_f)^N.
\end{equation}
We have
\begin{equation}
g'(p_f) = N[(2^N-1)(1-p_f)^{N-1} -(1+p_f)^{N-1}],
\end{equation}
which is equal to zero when $(1+p_f)^{N-1} = (2^N-1)(1-p_f)^{N-1}$. Inserting this into the expression for $g(p_f)$, we find the extremal value
\begin{equation}
g(p_f) = 2^N-2(1+p_f)^{N-1}
\end{equation}
which is strictly greater than 0 (unless $N=1$ 
or $p_f=1$, in which case we already know that inequality is satisfied in \eqref{eq:discineq}).
Since
\begin{equation}
g''(p_f)=-N(N-1)[(2^N-1)(1-p_f)^{N-2}+(1+p_f)^{N-2}]\le 0,
\end{equation}
the extremal value is a maximum for the function $g(p_f)$. It follows that the inequality in \eqref{eq:discineq} is not tight unless $N=1$, $p_f=0$ or $p_f=1$.}

\section{Conclusions}

We have derived unambiguous measurements for eliminating one and two out of the four possible two-qubit states in \eqref{eq:fourstates}, and established that for $N$-qubit sequences, where each qubit can be in one of two possible states, individual measurements on each qubit maximise the average number of states eliminated. That is, measurements in an entangled basis cannot exclude more states on average, but they can do as well as local measurements. Quantum state elimination has not been investigated as much as quantum state discrimination, but might be useful for applications in quantum communication and quantum information. As outlined in the introduction, excluding two states out of four has a connection to oblivious transfer and might for example be used for quantum key distribution. The result on the average number of states eliminated means, for example, that in any situation (such as a communication protocol) where what matters is to eliminate as many states as possible on average, a measurement in an entangled basis will not give any advantage.

There are many other scenarios one might consider.
Pusey et al.~\cite{PBR} also showed that for $N$ qubits, each in the state $|\pm\theta\rangle$, then no matter how close to zero $\theta$ is, it becomes possible to eliminate one of the $2^N$ $N$-qubit states with unit probability if $N$ is large enough. This requires a measurement in an entangled basis.
Another possibility is to have more than two possible states for each quantum system, meaning that each quantum system can also be more than two-dimensional. Some examples of this are considered in a related paper~\cite{Group}, where group theory is used to construct quantum state elimination measurements.

\acknowledgments
This work was supported by the UK Engineering and Physical Sciences Research Council (EPSRC) under EP/M013472/1 and  EP/L015110/1.


\begin{thebibliography}{00}

\bibitem{Stevebook} S. M. Barnett, {\em Quantum Information}, Oxford University Press, Oxford, pp 103-4 (2009).
\bibitem{PBR} M. Pusey, J. Barrett and T. Rudolph, {\em On the reality of the quantum state}, Nat. Phys. {\bf 8}, 475 (2012).
\bibitem{Caves2002} C. Caves, C. Fuchs and R. Schack, {\em Conditions for compatibility of quantum-state assignments}, Phys. Rev. A {\bf 66}, 062111 (2002).
\bibitem{SignaturesElim} R. J. Collins, R. Donaldson, V. Dunjko, P. Wallden, P. J. Clarke, E. Andersson, J. Jeffers, and G. S. Buller, {\em Realization of quantum digital signatures without the requirement of quantum memory}, Phys. Rev. Lett. {\bf 113}, 040502 (2014).
\bibitem{Bando2014} S. Bandyopadhyay, R. Jain, J.Oppenheim and C. Perry, {\em Conclusive exclusion of quantum states}, Phys. Rev. A {\bf 89}, 022336 (2014).
\bibitem{Wallden2014} P. Wallden, V. Dunjko and E. Andersson, {\em Minimum-cost measurements for quantum information}, J. Phys. A: Math. Theor. {\bf 47}, 125303 (2014).

\bibitem{Heinosaari1} \red{T. Heinosaari and O. Kerppo, {\em Antidistinguishability of Pure Quantum States}, J. Phys. A: Math. Theor. {\bf 51}, 365303 (2018).}

\bibitem{DiscReview} S. M. Barnett and S. Croke, {\em Quantum state discrimination}, Adv. Opt. Phot. {\bf 1}, 238-278 (2009).
\bibitem{Perry} C. Perry, R. Jain, and J. Oppenheim, {\em Communication tasks with infinite quantum-classical separation}, Phys. Rev. Lett. {\bf 115},  030504 (2015).

\bibitem{Heinosaari2} \red{ T. Heinosaari and O. Kerppo, {\em Communication of partial ignorance with qubits}, J. Phys. A: Math. Theor. {\bf 52}, 395301 (2019).}
\bibitem{Havlicek} \red{V. Havl\'{\i}\v{c}ek and J. Barrett,  {\em Simple Communication Complexity Separation from Quantum State Antidistinguishability}, arXiv:1911.01927 (2019).}

\bibitem{RyanOT} R. Amiri et al., to be submitted. 
\bibitem{Phoenix2000} S. J. D. Phoenix, S. M. Barnett, and A. Chefles, {\em Three-state quantum cryptography}, J. Mod. Opt. 47, 507 (2000).
\bibitem{Flatt2018} K. Flatt, S. Croke and S.M. Barnett, {\em Two-time state formalism for quantum eavesdropping}, Phys. Rev. A 98, 052339 (2018).

\bibitem{Ivan} I. D. Ivanovic, {\em How to differentiate between non-orthogonal states}, Phys. Lett. A \textbf{123}, 257 (1987).
\bibitem{Dieks} D. Dieks, {\em Overlap and distinguishability of quantum states}, Phys. Lett. A \textbf{126}, 303 (1988).
\bibitem{Peres} A. Peres, {\em How to differentiate between non-orthogonal states}, Phys. Lett. A \textbf{128}, 19 (1988). 
\bibitem{TBS} J. Crickmore et al., to be submitted. 
\bibitem{Stevecomp} S. M. Barnett, A. Chefles and I. Jex, {\em Comparison of two unknown pure quantum states}, 
Phys. Lett. A {\bf 307}, 189-195 (2003).
\bibitem{Comp1} A. Chefles, E. Andersson, and I. Jex, {\em Unambiguous comparison of the states of multiple quantum systems}, 
J. Phys. A: Math. Gen. {\bf 37}, 7315 (2004).
\bibitem{Comp2} I. Jex, E. Andersson and A. Chefles, {\em Comparing the states of many quantum systems}, 
J. Mod. Opt. {\bf 51}, 505 (2004).
\bibitem{Group} M. Hillery et al., to be submitted/arXiv:xxx (2019).

\end{thebibliography}
\end{document}